\documentclass[preprint,preprintnumbers,amsmath,amssymb,floatfix,showpacs,prl]{revtex4}

\usepackage{graphicx}

\begin{document}

\title{Nonstationary Regime of Random Lasers}
\author{Jonathan Andreasen}
\affiliation{
  Laboratoire de Physique de la Mati\`ere Condens\'ee, CNRS UMR 6622,
  Universit\'e de Nice-Sophia Antipolis,
  Parc Valrose, 06108, Nice Cedex 02, France
}
\author{Patrick Sebbah}
\affiliation{
  Laboratoire de Physique de la Mati\`ere Condens\'ee, CNRS UMR 6622,
  Universit\'e de Nice-Sophia Antipolis,
  Parc Valrose, 06108, Nice Cedex 02, France
}
\author{Christian Vanneste}
\affiliation{
  Laboratoire de Physique de la Mati\`ere Condens\'ee, CNRS UMR 6622,
  Universit\'e de Nice-Sophia Antipolis,
  Parc Valrose, 06108, Nice Cedex 02, France
}

\date{\today}

\begin{abstract}
A numerical study is presented of one-dimensional and two-dimensional random lasers as a 
function of the pumping rate above the threshold for lasing. 
Depending on the leakiness of the cavity modes, 
we observe that the stationary lasing regime becomes unstable above a second threshold.
Coherent instabilities are observed as self pulsation of the total
output intensity, population inversion, and polarization.

\end{abstract}


\maketitle

Since their prediction by Lethokov \cite{letokhov68a}, 
random lasers have been the subject of numerous studies. 
After the first experiments that started at the end of the 80's \cite{markushev}, 
the observation of narrow lines in the laser spectra at the end of the 90's \cite{cao98,cao99,frolov99a}
led to a several year's debate about the nature of lasing modes in such open systems without a cavity. 
Important progress in the understanding of random lasers has been recently made in theoretical and 
numerical studies that show some correspondence exists between lasing modes and resonances 
of the passive system without gain \cite{vannestePRL01,souk02,VannestePRL,tureciSci,tureciN09,andreasennu}.
A recent review devoted to the first lasing mode at threshold \cite{review}
shows how this correspondence depends on the openness of the system. 
The interaction of lasing modes in the multimode regime of a random laser is not so well known. 
Experimentally, it has been specifically investigated in a few studies at the beginning of the 2000's 
\cite{soukoulisPRB02,caoPRB03,jiangPRB04}.
Theoretically, only recently the \textit{ab initio} laser theory \cite{tureciN09} was able 
to make precise predictions about the multimode lasing spectrum above threshold when mode competition takes place. 
These various works deal with different aspects of multimode random lasing. 
The first series of works concerns the dynamic response of random lasers using picosecond optical pumping. 
These random lasers are in a transient regime rather than a steady regime. 
The \textit{ab initio} laser theory is devoted to the stationary regime of lasing in open systems under steady pumping.
It relies on the assumption that the atomic population inversion is time-independent and that there exists a 
steady-state multiperiodic solution of the laser field and polarization of the atomic medium. 
An important issue remains unaddressed:
a stationary regime may not always exist under steady pumping of a random laser.
Though different kinds of instabilities and nonstationary regimes are known to take place in conventional lasers, 
they have never been reported in random lasers.

In this paper, we numerically investigate one-dimensional (1D) and two-dimensional (2D) random lasers 
using steady external pumping by progressively increasing the pump intensity. 
After observing stationary single-mode lasing and multimode lasing, 
we find that the stationary regime can become unstable for sufficiently large values of the pumping rate. 
This instability occurs regularly in 1D lasers and systematically in 2D lasers. 
It appears as time oscillations of the output intensity, 
atomic population inversion, and polarization. 
We show that the instability is associated with short resonance lifetimes of the random systems.
It is found to be related to Rabi oscillations and of the same type as instabilities discovered earlier
in conventional lasers \cite{hakenPL75,riskenJAP68,grahamZP68}.

The 1D random structures we consider are composed of $41$ layers.
Dielectric material with optical index $n_1=1.25$ separated by air gaps
($n_2=1$) results in a spatially modulated index $n(x)$.
Outside the random medium, the index is 1. 
The system is randomized by specifying thicknesses for each
layer as $d_{1,2} = \left<d_{1,2}\right>(1+\eta\zeta)$, where
$\left<d_1\right>=100$ nm and $\left<d_2\right>=200$ nm are the average
thicknesses of the layers, $\eta = 0.9$ represents the degree of randomness,
and $\zeta$ is a random number in (-1,1).
The length of the random structure $L$ is normalized to
$\left<L\right>=6.1$ $\mu$m.
These parameters give a localization length $\xi \approx 11$ $\mu$m. 

The 2D random structures we consider are
of size $L^2=5\times 5$ $\mu$m$^2$ made of circular dielectric particles with radius
$r=60$ nm, optical index $n_1=1.25$, and surface filling
fraction $\Phi = 40$\%, which are randomly distributed in a
background medium of index $n_2=1$. 
Outside the random medium, the index is 1. 
The scattering mean free path $\ell_s\approx 2$ $\mu$m
and the localization length $\xi \approx 12$ $\mu$m.
The system is in the weakly scattering regime--intermediate between ballistic and diffusive. 

\begin{figure}
\begin{center}
  \includegraphics[width=4.25cm]{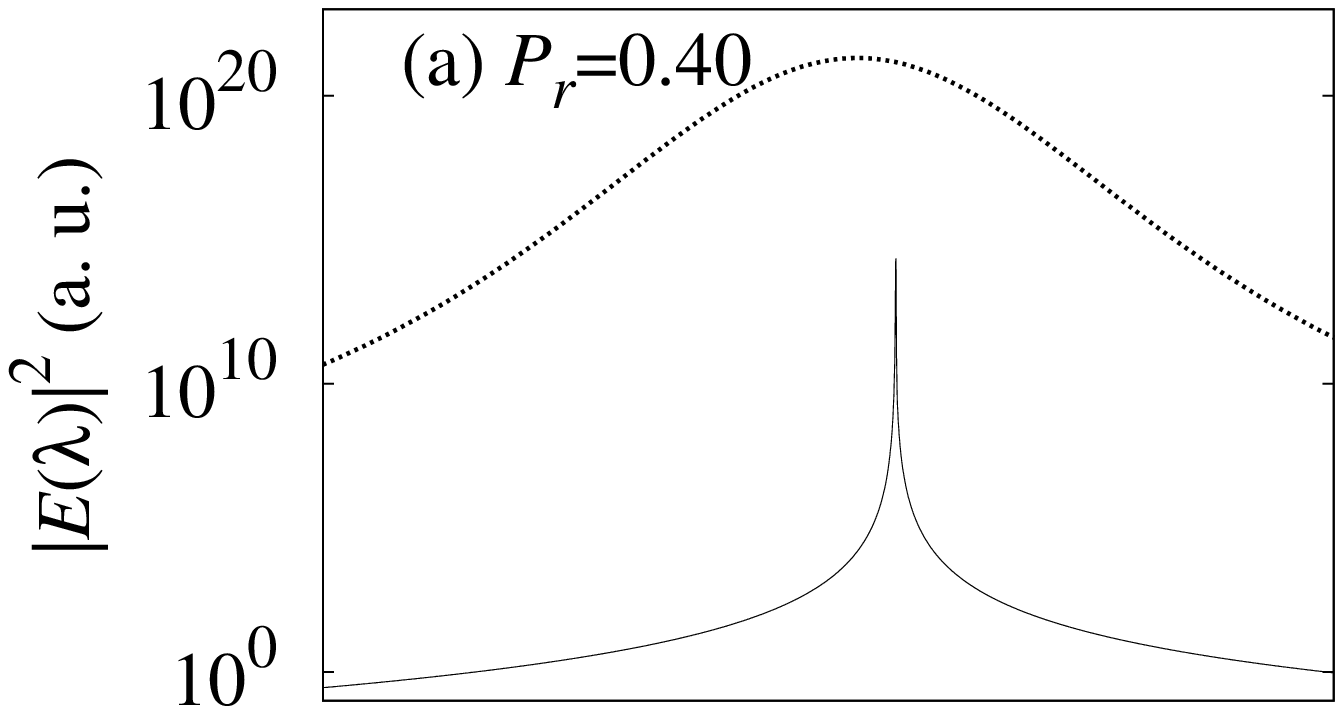}
  \includegraphics[width=4.25cm]{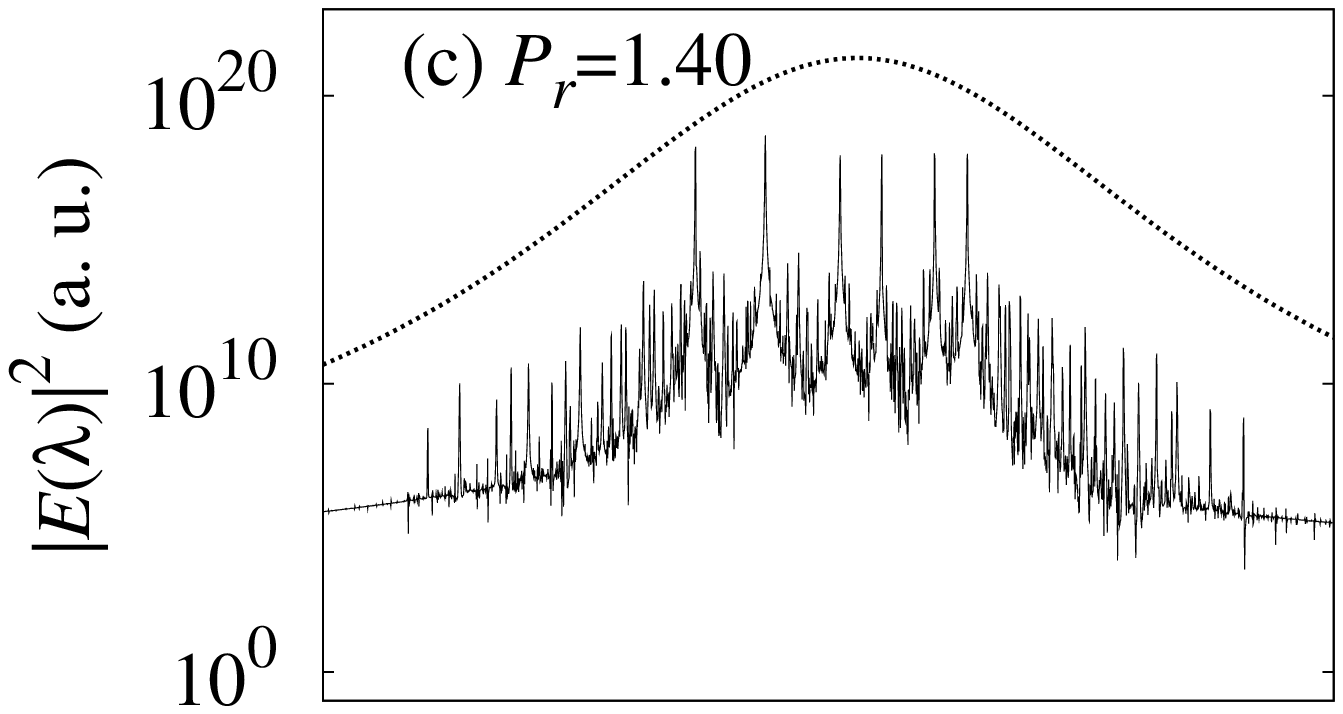}\\
  \includegraphics[width=4.25cm]{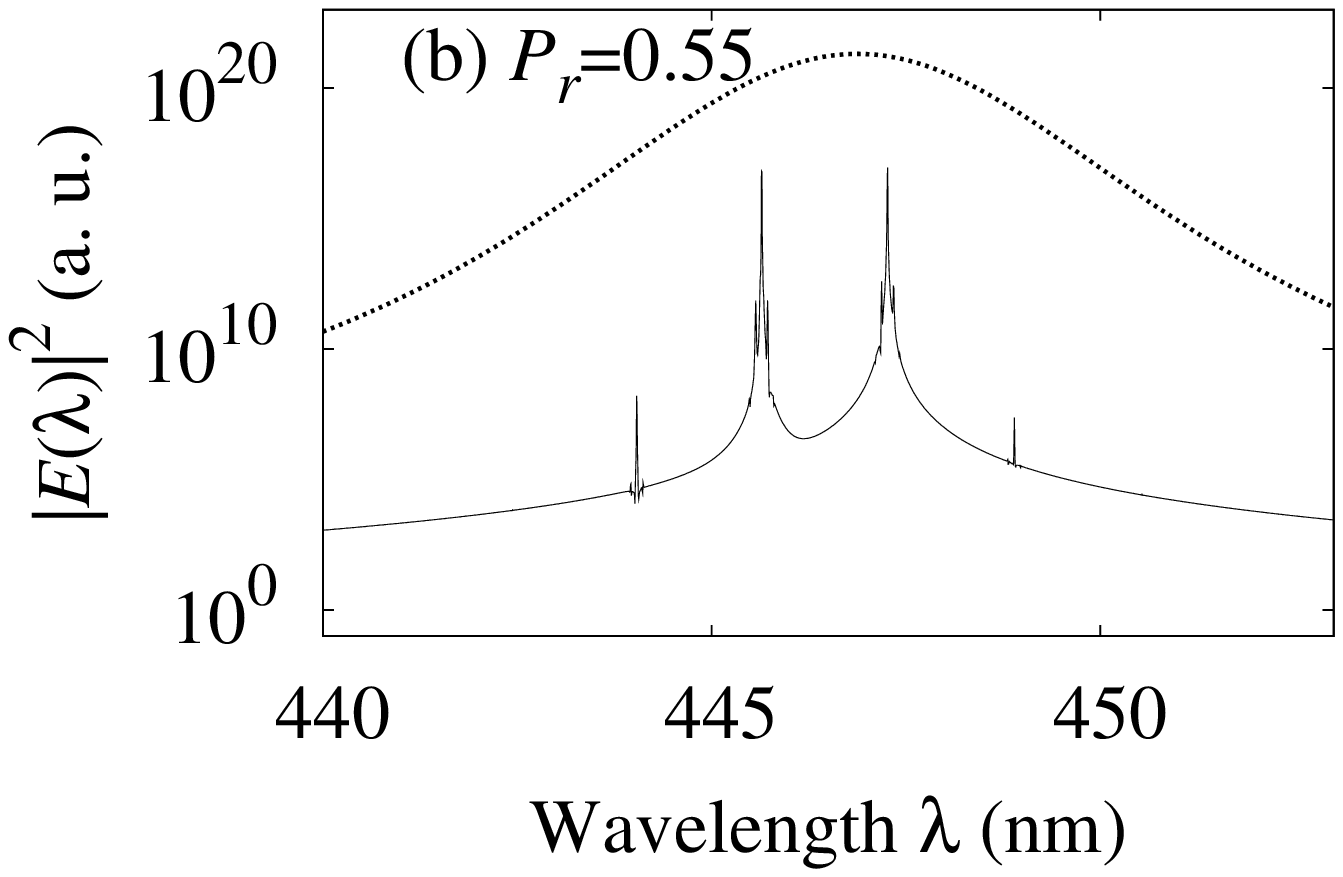}
  \includegraphics[width=4.25cm]{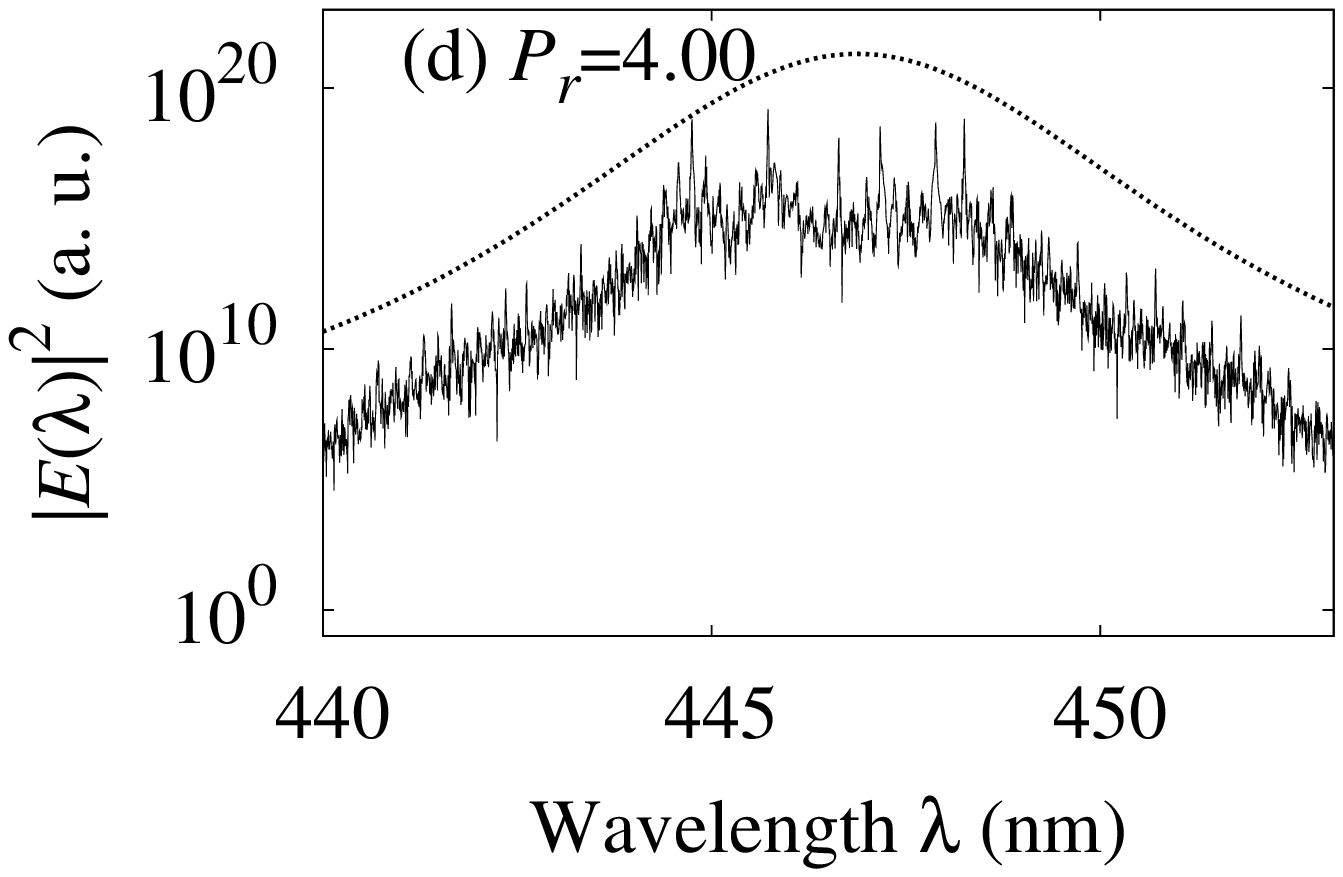}\\
  \caption{\label{fig:fig1}
    Emission spectra $|E(\lambda)|^2$ for a 2D random system.
    The values of the pumping rate $P_r$ are written in each panel
    in units ns$^{-1}$.
    The gain curve (dotted line) is overlaid.
    The two side peaks in (b) with smaller amplitudes (by 9 orders of magnitude) 
    are due to four-wave mixing. 
  }
\end{center}
\end{figure}

The background medium (air) is chosen as the active part of the system and is modeled as a
four-level atomic system. 
We describe the time evolution of the fields by
Maxwell's equations including a polarization term due to the atomic population inversion.
Maxwell's equations are solved using the finite-difference time-domain (FDTD) method 
and C-PML absorbing boundary conditions in order to model the open system \cite{tafl05}.
The corresponding equations are identical to those used in \cite{sebbah02},
where $T_1=100$ ps and $T_2=20$ fs so that the gain curve centered at 
$\lambda_a=446.9$ nm has a spectral width $\Delta\lambda_a=11$ nm. 
The control parameter is the pumping rate $P_r$ at which an external mechanism transfers 
the atoms from the ground-state level $0$ to the upper level $3$ of the four-level system. 

Figure \ref{fig:fig1} shows the emission spectra $|E(\lambda)|^2$
from a 2D random laser with increasing pumping rates.
Just above the lasing threshold at $P_r=0.40$ ns$^{-1}$, 
a single lasing peak is observed in the emission spectrum [Fig. \ref{fig:fig1}(a)]. 
The lasing wavelength $\lambda=447.37$ nm is close to but not coincident with
the gain center wavelength $\lambda_a$.
This is indicative of lasing associated with a resonance of the random system \cite{review}.
The multimode regime is reached by $P_r=0.55$ ns$^{-1}$,
with a second lasing peak appearing at $\lambda=445.65$ nm [Fig.\ref{fig:fig1}(b)].
We find a different spatial distribution of intensity at the second lasing wavelength indicating
this lasing mode is associated with a different resonance of the random system.
With the pumping rate increased to $P_r=1.40$ ns$^{-1}$ [Fig.\ref{fig:fig1}(c)],
a huge number of fine spectral features appears
while the original lasing peaks are still observed.
Such behavior persists for higher pumping rates [Fig.\ref{fig:fig1}(d) where $P_r=4.00$ ns$^{-1}$].

\begin{figure}
\begin{center}
  \includegraphics[width=4.25cm]{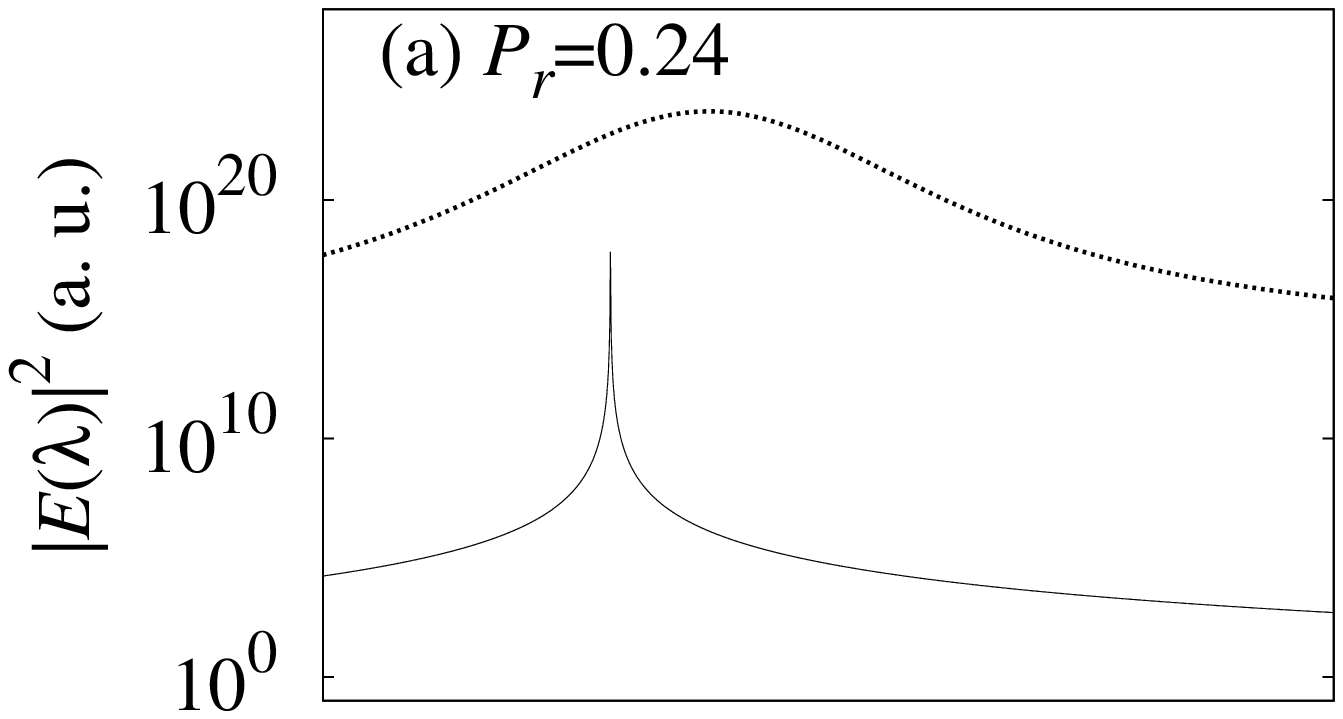}
  \includegraphics[width=4.25cm]{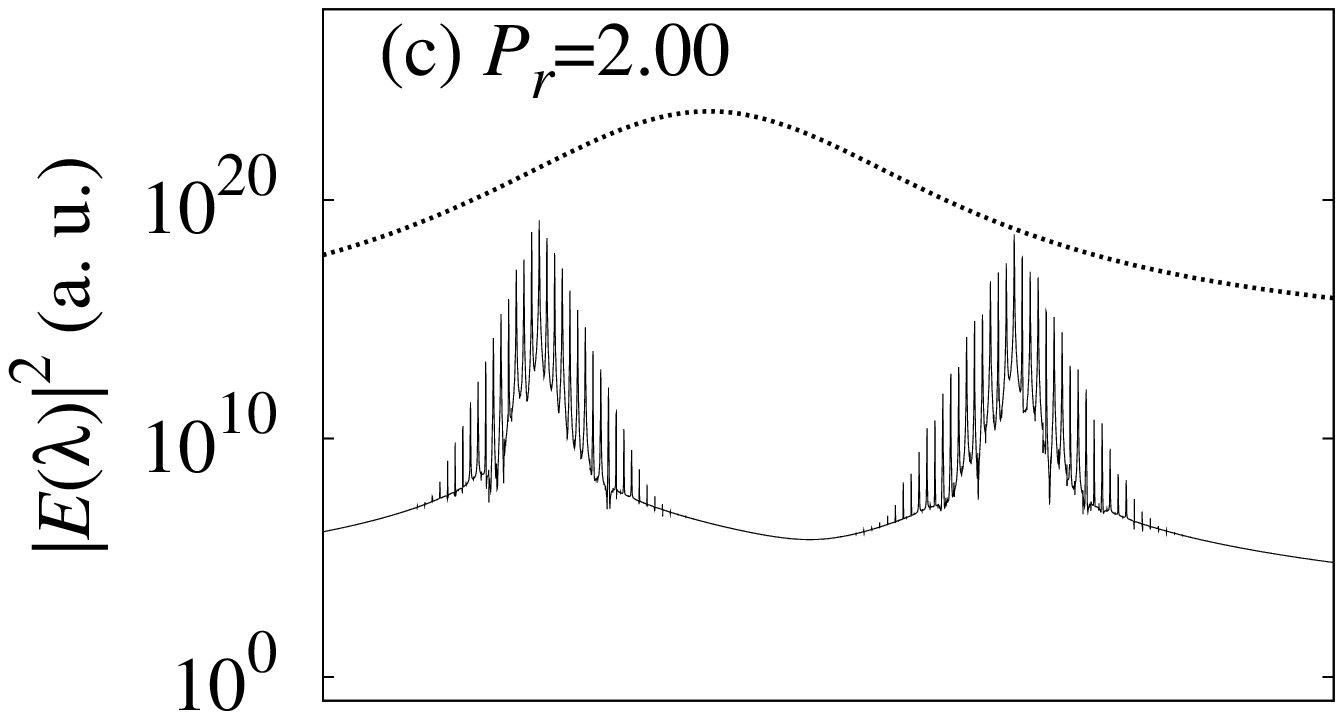}\\
  \includegraphics[width=4.25cm]{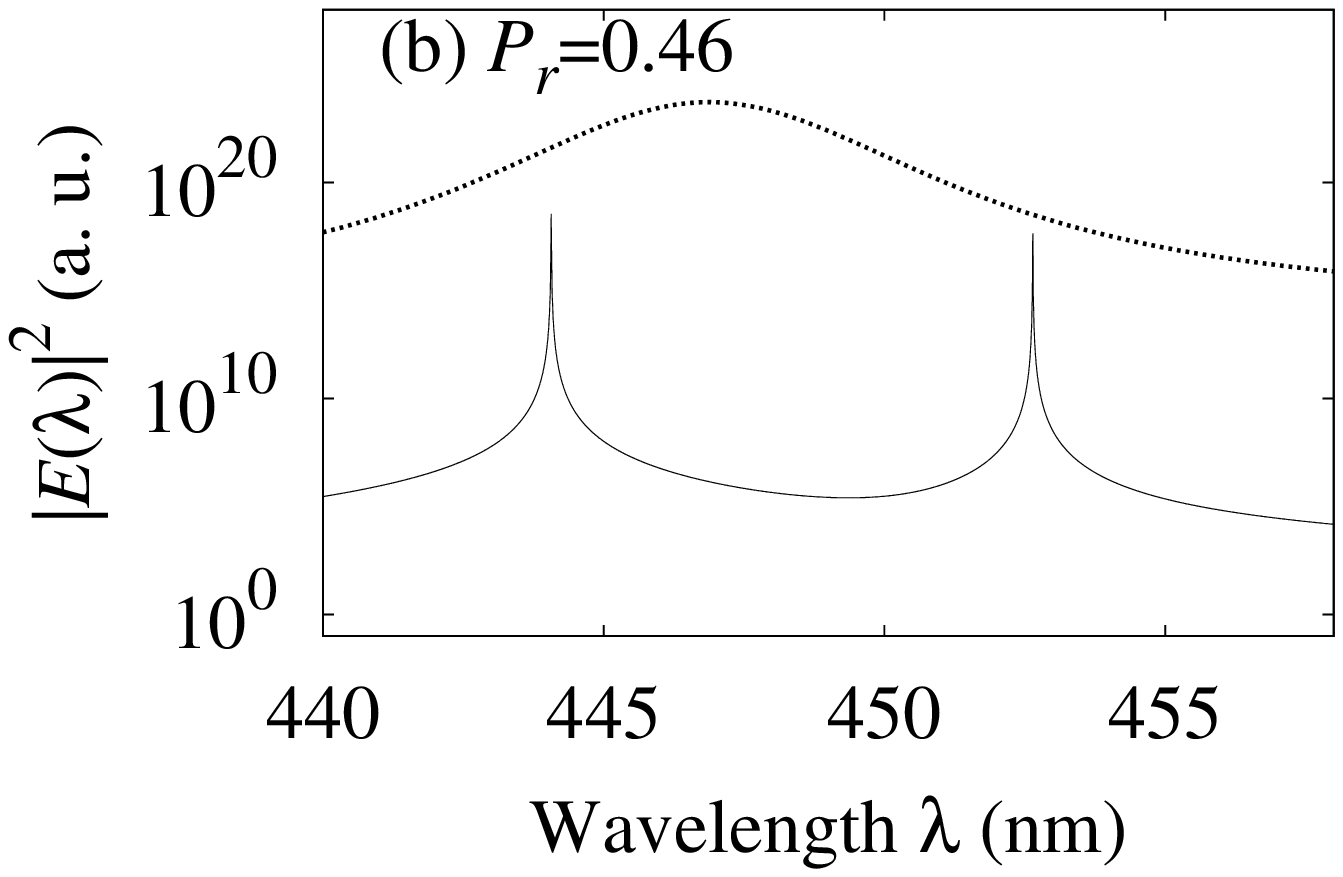}
  \includegraphics[width=4.25cm]{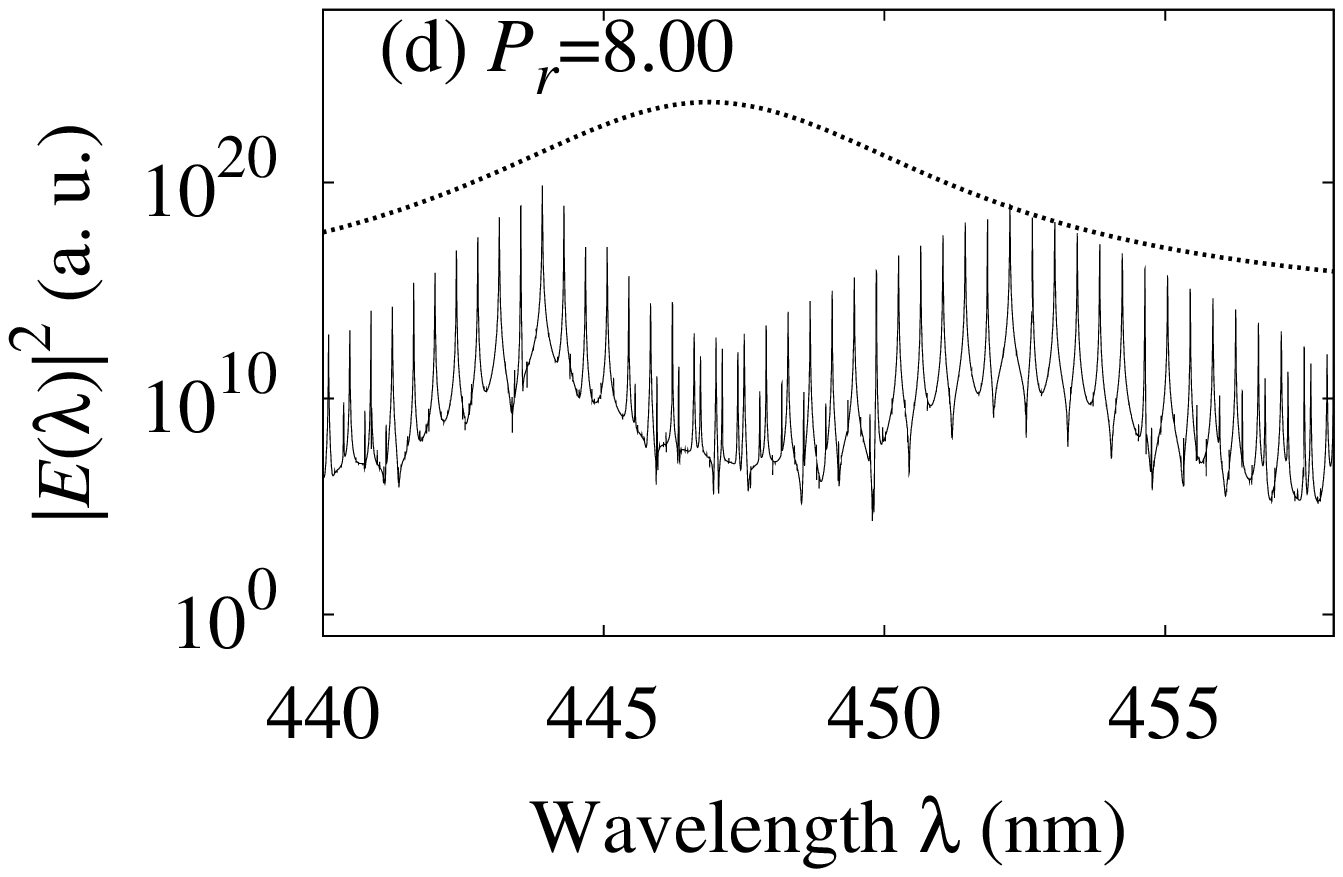}\\
  \caption{\label{fig:fig2} 
    Emission spectra $|E(\lambda)|^2$ for a 1D random system.
    The values of the pumping rate $P_r$ are written in each panel
    in units ns$^{-1}$.
    The gain curve (dotted line) is overlaid.
  }
\end{center}
\end{figure}

For a more thorough study of this high-pumping regime without
the complications caused by a large density of lasing modes,
we switch to a study of 1D random lasers.
Figure \ref{fig:fig2} shows the emission spectra $|E(\lambda)|^2$
with increasing pumping rates.
Just above the lasing threshold at $P_r=0.24$ ns$^{-1}$,
a single lasing peak is observed in the emission spectrum [Fig. \ref{fig:fig2}(a)].
The lasing wavelength $\lambda=445.12$ nm is close to but not coincident with
the gain center wavelength $\lambda_a=446.90$ nm.
We verified this lasing mode's association with a resonance of the random system \cite{born75}
by comparing their intensity distributions.
The multimode regime is reached by $P_r=0.46$ ns$^{-1}$,
with a second lasing peak appearing at $\lambda=452.64$ nm [Fig.\ref{fig:fig2}(b)].
The second lasing mode is associated with a second resonance (verified by comparisons of intensity distributions).

With the pumping rate increased to $P_r=2.00$ ns$^{-1}$ [Fig.\ref{fig:fig2}(c)],
a huge number of fine spectral features appear as in the 2D case.
Numerous sidebands are generated on either side of the two lasing wavelengths.
Such behavior persists for higher pumping rates,
but the frequency spacing between the fine spectral features increases [Fig.\ref{fig:fig2}(d) where $P_r=8.00$ ns$^{-1}$].
We emphasize that the sideband peaks cannot correspond to system resonances because their frequency spacing
is much smaller than the intermode spacing shown clearly in Fig. \ref{fig:fig2}(b).

To investigate the cause of this spectral behavior, we examine the dynamics of the system.
The output intensity $|E(t)|^2$ is sampled on one side of the random system
and averaged over each optical period $T_a=\lambda_a/c$.
The population inversion $\Delta N$ is averaged spatially
\begin{equation}
  \Delta N(t) = \int_0^{\tilde{L}} \left[ N_2(x,t) - N_1(x,t) \right] dx/\tilde{L}N_a,\label{eq:eq1}
\end{equation}
where $N_a$ is the total density of atoms and the integration is over the gain domain $\tilde{L}$.
The fluctuations of the population inversion are measured as the
temporal standard deviation
\begin{equation}
  \sigma^2=\int_{t_1}^{t_2} \left[\Delta N(t) - \left<\Delta N(t)\right> \right]^2 dt/ (t_2-t_1).\label{eq:eq2}
\end{equation}

\begin{figure}
\begin{center}
  \includegraphics[width=6cm]{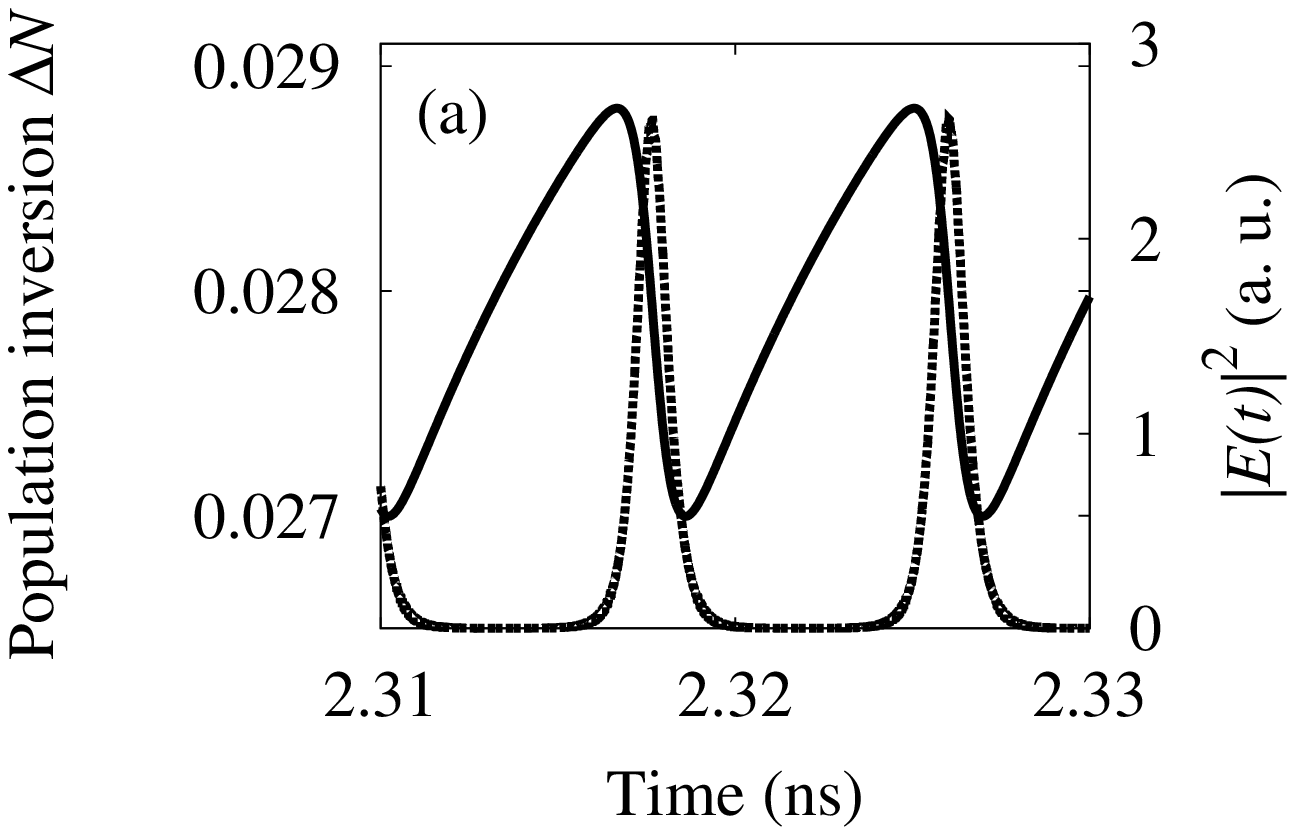}\\
  \includegraphics[width=6cm]{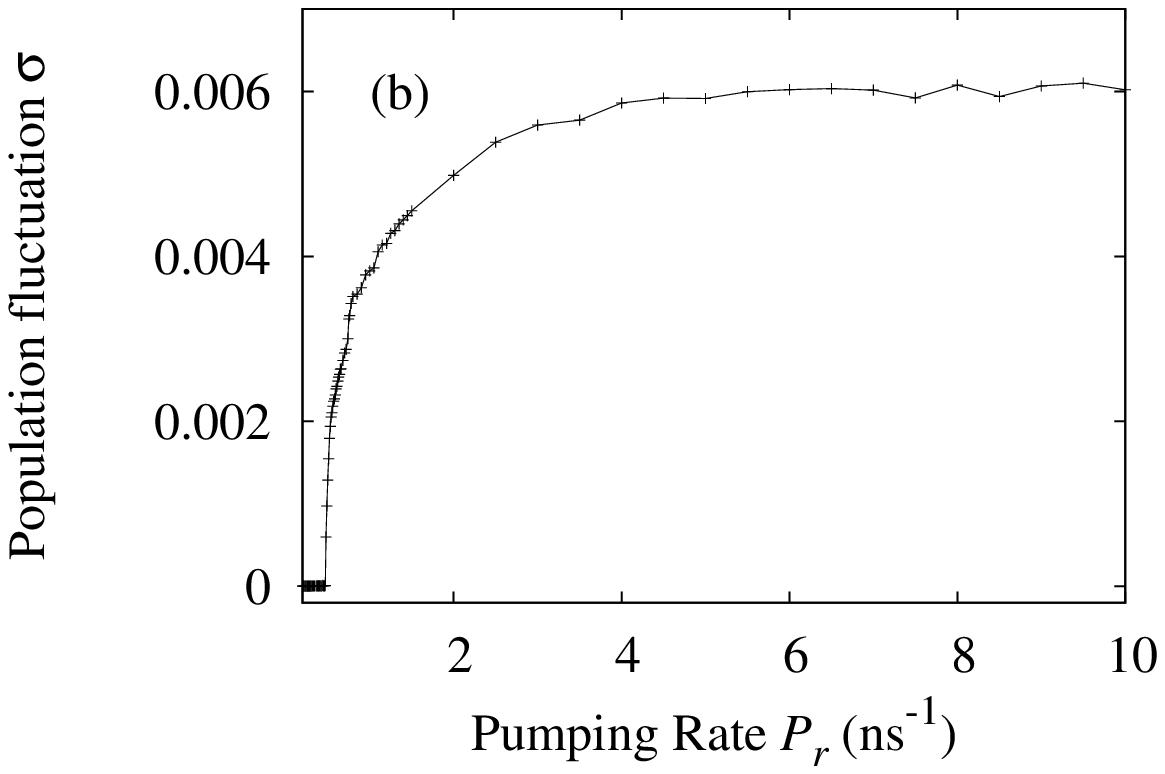}\\
  \caption{\label{fig:fig3}
    (a) Population inversion (solid line) and intensity $|E(t)|^2$ (dotted line)
    for a 1D random system with $P_r= 0.48$ ns$^{-1}$.
    (b) Temporal standard deviation $\sigma$ of the population inversion
    vs. pumping rate $P_r$.
  }
\end{center}
\end{figure}

Dynamic behavior for the smallest pumping rate at which the spectral splitting occurs
$P_r=0.48$ ns$^{-1}$ (2 times the lasing threshold) is shown in Fig. \ref{fig:fig3}(a). 
A time interval late in the simulation is chosen to avoid the transient regime where
relaxation oscillations (with a decay time $\tau_{ro} \approx 100$ ps) occur.
The population inversion clearly oscillates resulting in a series of optical pulses. 
Below the instability threshold, in the multimode regime, the polarization and population 
inversion also fluctuate in time due to mode beating.
However, these fluctuations are small.
In contrast, at these larger pumping rates, temporal fluctuations increase dramatically putting
the system in a highly nonstationary regime. 
Figure \ref{fig:fig3}(b) shows the temporal standard deviation $\sigma$ of $\Delta N(t)$ 
with increasing $P_r$,
where the time interval was chosen as $t_1=333$ ps and $t_2=667$ ps
to avoid the transient regime.
In this case, $\sigma$ increases by 2 orders of magnitude from $P_r=0.47$ ns$^{-1}$
to $0.48$ ns$^{-1}$ and keeps increasing until $P_r=4.00$ ns$^{-1}$. 
We find these fluctuations to be coherent in nature, involving the polarization of the gain medium which
is observed to fluctuate on the same time scale as the population inversion and intensity seen in Fig. \ref{fig:fig3}(a).

\begin{figure}
\begin{center}
  \includegraphics[width=4.25cm]{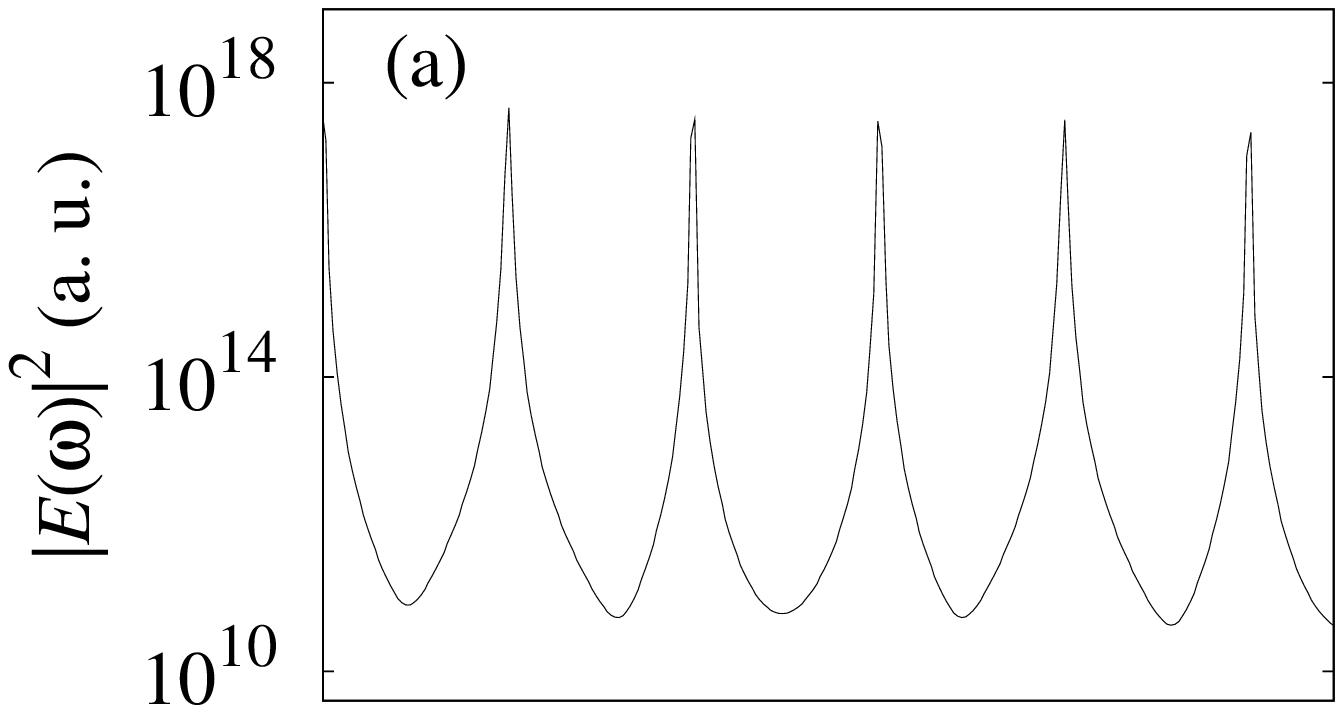}
  \includegraphics[width=4.25cm]{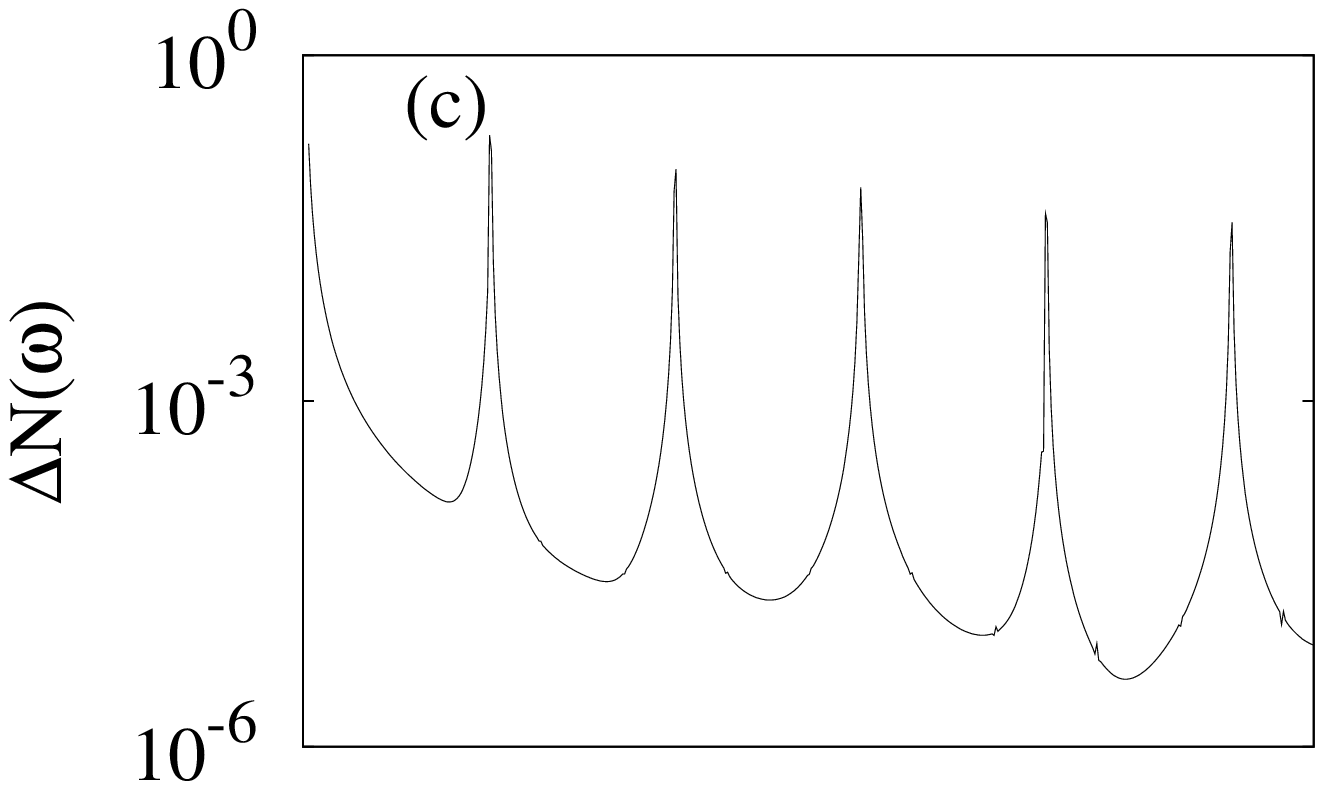}\\
  \includegraphics[width=4.25cm]{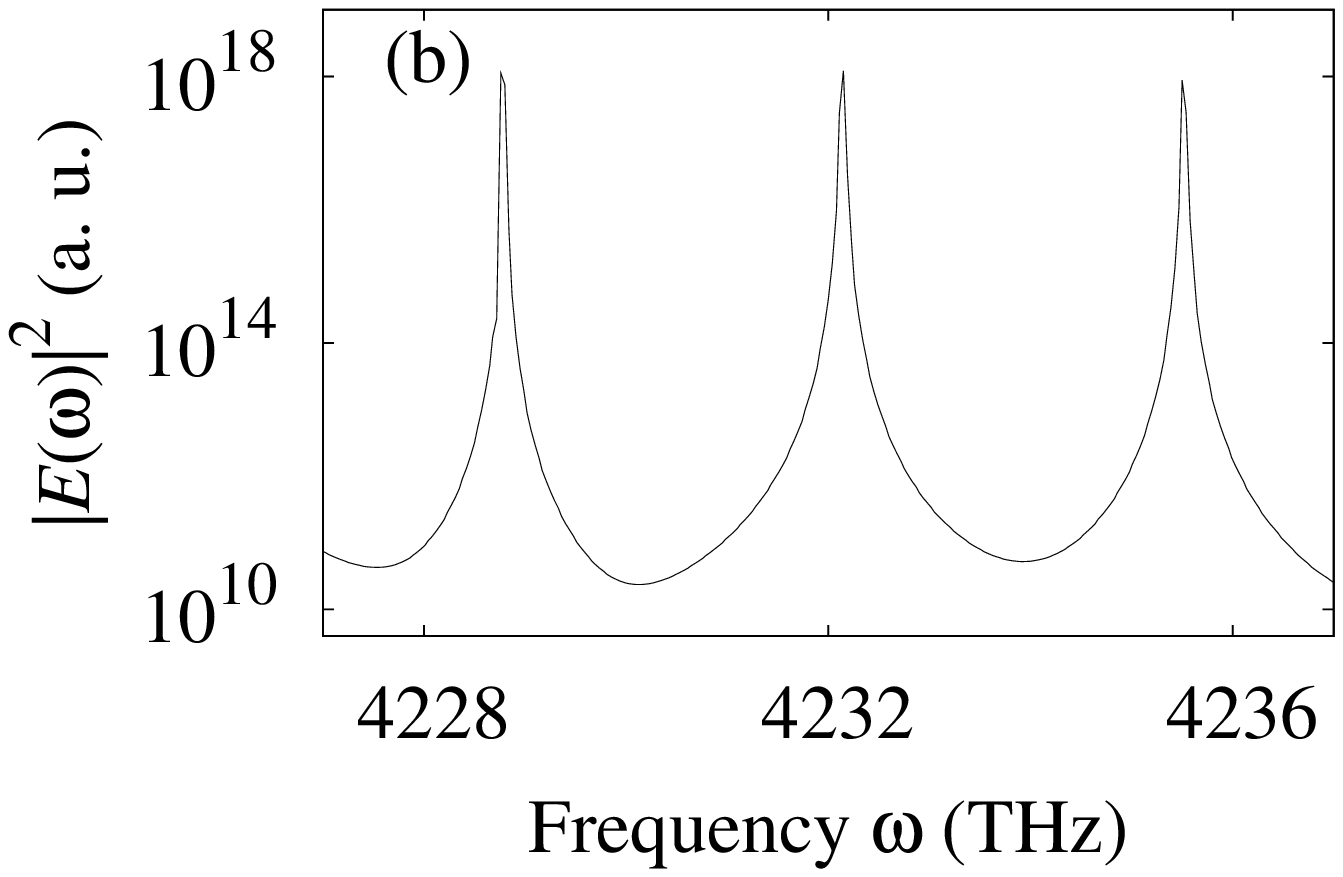}
  \includegraphics[width=4.25cm]{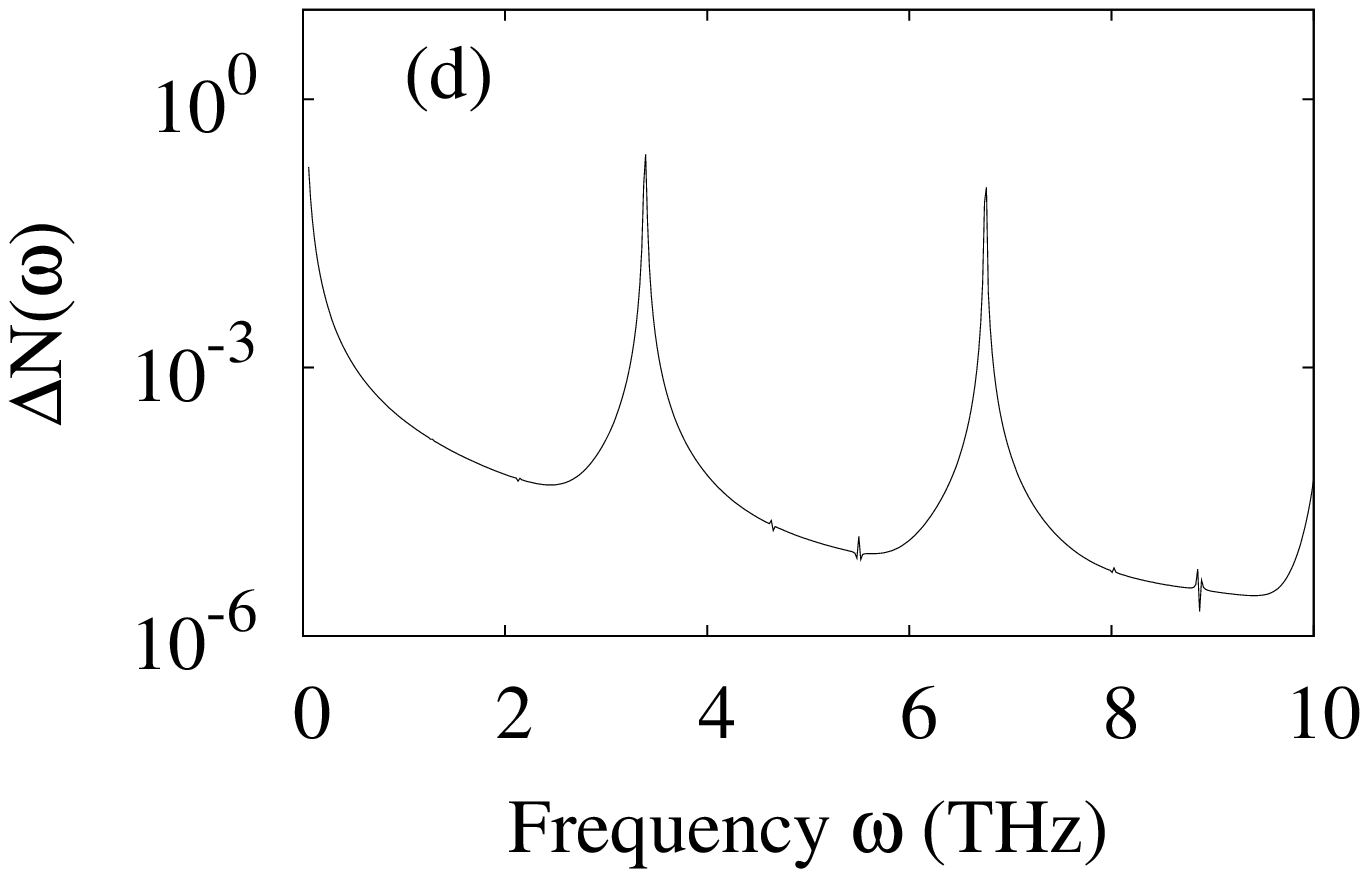}\\
  \caption{\label{fig:fig4} 
    Emission spectra $|E(\omega)|^2$ (a--b)
    and population inversion $\Delta N(\omega)$ (c--d) with
    $P_r=4.00$ ns$^{-1}$ (a,c) and $P_r=8.00$ ns$^{-1}$ (b,d) for a 1D random system.
    $\omega_p$ is the smallest (nonzero) frequency peak in $\Delta N(\omega)$ and
    corresponds to the frequency spacing in the emission spectrum.
  }
\end{center}
\end{figure}

Spectral behavior of laser emission in a smaller frequency range is shown in Figs. \ref{fig:fig4}(a) and \ref{fig:fig4}(b). 
A frequency comb is evident with regularly spaced peaks.
The frequency spacing between peaks increases from $\Delta\omega=1.8$ THz to $3.4$ THz 
as the pumping rate increases from $P_r=4.00$ ns$^{-1}$ to $8.00$ ns$^{-1}$.
Figures \ref{fig:fig4}(c) and \ref{fig:fig4}(d) show the population inversion spectrum
$\Delta N(\omega)$ [calculated from $\Delta N(t)$ in Eq. (\ref{eq:eq1})]. 
The first and largest non-zero peaks in $\Delta N(\omega)$ for the two pumping rates
are at $\omega_p=1.8$ THz and $3.4$ THz. 
Thus, $\omega_p$ corresponds precisely to the frequency spacing $\Delta\omega$ of the laser emission 
for both pumping rates. 
We verified that these two values equal each other for intermediate pumping rates as well.
The additional higher-frequency peaks in $\Delta N(\omega)$ are harmonics related to
the sharp sawtooth behavior of $\Delta N(t)$ seen in Fig. \ref{fig:fig3}(a).

The Rabi frequency $\Omega$ in a random laser is not a well-defined quantity since
the field $E(x,t)$ varies spatially. 
In this nonstationary regime, all relevant quantities fluctuate in time.
For a fixed pumping rate, however, their values averaged over large enough time windows 
become nearly constant eventually.
Focusing on a late temporal region and avoiding the transient regime, 
we calculate the spatiotemporally averaged intracavity field amplitude $E_a$
and estimate the Rabi frequency as
\begin{equation}
  \Omega = \gamma E_a/\hbar,\label{eq:eq3}
\end{equation}
where $\gamma = \sqrt{{3\lambda_a^3\hbar\epsilon_0}/ {8\pi^2T_1}}= 1.8 \times 10^{-28}$ C$\cdot$m
is the atomic dipole coupling term.

\begin{figure}
\begin{center}
  \includegraphics[width=6cm]{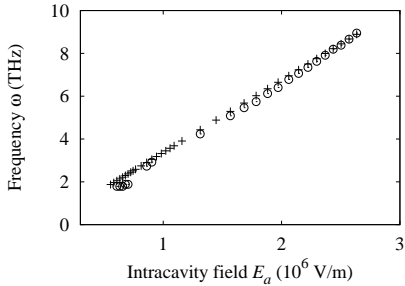}
  \caption{\label{fig:fig5}
    Estimated Rabi frequency $\Omega$ (crosses) and the population inversion frequency
    $\omega_p$ (circles) for a 1D random system vs. the intracavity field $E_a$. 
  }
\end{center}
\end{figure}

Figure \ref{fig:fig5} compares $\omega_p$ to $\Omega$.
We find good agreement for a wide range of pumping rates
(from 2--40 times the lasing threshold).
This agreement shows the spectral splitting and population inversion oscillations
are related to the Rabi frequency.

It is known that coherent instabilites in homogeneously broadened lasers \cite{narducci88} take place in different situations,
such as the Lorenz-Haken (LH) \cite{lorenzJAS63,hakenPL75}
or the Risken-Nummedal-Graham-Haken (RNGH) \cite{riskenJAP68,grahamZP68} instability.
Coherent instabilities can arise when the mode lifetime is short enough so that atomic relaxations
do not destroy the coherent Rabi oscillations, i.e., $\tau < (1/T_1 + 1/T_2)^{-1}$.
In this so-called bad-cavity limit,
sideband peaks appear in the spectra separated by the Rabi frequency due to gain provided by the Rabi oscillations.
For the resonances associated with the two lasing modes in Fig. \ref{fig:fig2},
the lifetimes are $\tau=11$ fs and $\tau=16$ fs, respectively, thus satisfying the condition. 
Another condition on such instabilities is that the Rabi frequency must be greater than the inverse atomic relaxation times
so the coherent oscillations are not destroyed.
This means the intensity and therefore the pumping rate must be sufficiently large.
The instability threshold observed here is roughly twice the lasing threshold.
Many factors may contribute to the precise threshold value, such as 
a saturable absorber \cite{wangPRAR07,gordonPRA08}
or spatially varying intensity \cite{urchueguiaPRA00,chenkosolJOSAB09}.

We examined the behavior of different realizations of random structures in 1D and 2D. 
We find that when the instability manifests itself in 1D, there is rather good agreement
between $\Delta\omega$, $\omega_p$, and $\Omega$ as in Fig. \ref{fig:fig5}.
Differences between these quantities from realization to realization are likely due to the fact that 
the correct value of the field to be used in Eq. (\ref{eq:eq3}) is not exactly known.
Nevertheless, we find that $\omega_p$ scales linearly with $E_a$ along with the Rabi frequency. 
Furthermore, the condition on $\tau$ is always satisfied and the instability threshold 
is always roughly twice the lasing threshold. 
When the condition on $\tau$ is not satisfied, the instability does not occur.
In 2D, we consistently find spectral behavior as that shown in Fig. \ref{fig:fig1}
and have not found a case in which the instability did not manifest itself.
This is not surprising since mode lifetimes are likely shorter in 2D than in 1D
for the systems we considered. 
Due to the small density of states, the results in 1D seem to be related to the
single mode LH instability \cite{lorenzJAS63,hakenPL75} as opposed to the multimode RNGH instability \cite{riskenJAP68,grahamZP68}.
Because the density of states is larger in 2D than in 1D, the analysis of the 2D case is more complicated. 
Though we could find some behavior close to the 1D cases, most systems exhibit an instability where 
the spectrum is complicated by the presence of simultaneous lasing resonances and four-wave-mixing peaks. 
Hence, it is difficult to distinguish between the multimode or single mode instability as in 1D. 
Further investigation of 2D systems is needed.

In conclusion, when increasing the pumping rate above the lasing threshold, we found that the stationary 
regime of random lasers is not stable and gives rise to a coherent instability due
to very short lifetimes of the system resonances. 
Only in the uncommon experimental situation of modes with long lifetimes, such as those in the localization regime, 
should the instability not be observed.
The bad-cavity limit, and hence the nonstationary regime, is likely to be reached in experiments because:
(i) random lasers are open systems with strong leakage and
(ii) atomic lifetimes are typically longer than those we used in this paper.
After recent progress in the theoretical understanding of random lasers in the stationary regime \cite{review},
this work is a step toward understanding the rich physics of such lasers which combine
random systems, complex light-matter interaction, and nonlinear dynamic behavior.

\begin{acknowledgments}
  We thank H. Cao for stimulating discussions.
  This work was supported by the ANR under Grant No. ANR-08-BLAN-0302-01, 
  the PACA region, the CG06, and the Groupement de Recherche 3219 MesoImage.  
  JA acknowledges support from the Chateaubriand Fellowship. 
  This work was performed using HPC resources from GENCI-CINES (Grant 2010-99660). 
\end{acknowledgments}



\end{document}